\documentclass[sigconf]{acmart}

\AtBeginDocument{%
  \providecommand\BibTeX{{%
    \normalfont B\kern-0.5em{\scshape i\kern-0.25em b}\kern-0.8em\TeX}}}

\usepackage{hyperref}
\usepackage{color}
\usepackage{soul}
\usepackage{multirow}
\usepackage{graphicx}
\usepackage{caption}
\usepackage{subcaption}
\usepackage{textcomp}
\usepackage{algorithm, algorithmic}

\copyrightyear{2021}
\acmYear{2021}
\setcopyright{acmcopyright}
\acmConference[ISEC 2021]{14th Innovations in Software Engineering Conference (formerly known as India Software Engineering Conference)}{February 25--27, 2021}{Bhubaneswar, Odisha, India}
\acmBooktitle{14th Innovations in Software Engineering Conference (formerly known as India Software Engineering Conference) (ISEC 2021), February 25--27, 2021, Bhubaneswar, Odisha, India}
\acmPrice{15.00}
\acmDOI{10.1145/3452383.3452393}
\acmISBN{978-1-4503-9046-0/21/02}

\begin{document}

\title{Using Paragraph Vectors to improve our existing code review assisting tool-CRUSO}

\author{Ritu Kapur}
\authornote{Both authors contributed equally to this research.}
\email{dev.ritu.kapur@gmail.com}
\orcid{0001-7112-0630}
\author{Balwinder Sodhi}
\authornotemark[1]
\email{sodhi@iitrpr.ac.in}
\orcid{0001-7137-5529}
\affiliation{%
  \institution{Indian Institute of Technology Ropar}
    \email{poojith.19csz0006@iitrpr.ac.in }
  \streetaddress{101 Ramanujan Block, Bara Phool}
  \city{Rupnagar}
  \state{Punjab}
  \country{India}
  \postcode{140001}
}

\author{Poojith U Rao}
\affiliation{%
  \institution{Indian Institute of Technology Ropar}
    \email{poojith.19csz0006@iitrpr.ac.in }
  \streetaddress{101 Ramanujan Block, Bara Phool}
  \city{Roopnagar}
  \state{Punjab}
  \country{India}
  \postcode{140001}
}

\author{Shipra Sharma}
\affiliation{%
  \institution{Indian Institute of Technology Ropar}
  \streetaddress{101 Ramanujan Block, Bara Phool}
  \email{shipra@iitrpr.ac.in}
  \city{Roopnagar}
  \state{Punjab}
  \country{India}
  \postcode{140001}
}

\renewcommand{\shortauthors}{Ritu et al.}

\begin{abstract}
 Code reviews are one of the effective methods to estimate defectiveness in source code. However, the existing methods are dependent on experts or inefficient. 
In this paper, we improve the performance (in terms of speed and memory usage) of our existing code review assisting tool--CRUSO.
The central idea of the approach is to estimate the defectiveness for an input source code by using the defectiveness score of similar code fragments present in various StackOverflow (SO)  posts. 

The significant contributions of our paper are i) \emph{SOpostsDB}: a dataset containing the PVA vectors and the SO posts information, 
ii) \emph{CRUSO-P}: a code review assisting system based on PVA models trained on \emph{SOpostsDB}.
For a given input source code, CRUSO-P labels it as \{\texttt{Likely to be defective, Unlikely to be defective, Unpredictable}\}. To develop CRUSO-P, we processed >3 million SO posts and 188200+ GitHub source files. CRUSO-P is designed to work with source code written in the popular programming languages \{C, C\#, Java, JavaScript, and Python\}. 

CRUSO-P outperforms CRUSO with an improvement of 97.82\% in response time and a storage reduction of 99.15\%. CRUSO-P achieves the highest mean accuracy score of 99.6\% when tested with the C programming language, thus achieving an improvement of 5.6\% over the existing method.  
\end{abstract}



\maketitle
\section{Introduction}
Code reviews play a significant role in detecting potential defects that remain undiscovered through the software testing process. Some such examples include memory leaks, buffer overflows, and scalability issues. However, the existing methods to perform the code reviews are dependent on subject-matter experts (SMEs) and being significantly time-consuming \cite{codereview2019wastetime}. Therefore, we worked on \textbf{improving the performance of our existing code review assisting tool}-- CRUSO \cite{Sodhi2019UsingSC}.

For a given source code $c$, CRUSO performs the following steps: 
\begin{enumerate}
    \item Identifies the set of StackOverflow (SO)\footnote{\url{https://stackoverflow.com}} posts $P$ such that each $p\in P$ contains source code fragment(s), which sufficiently resemble $c$.
    \item Determines the likelihood of $c$ being defective by considering all $p \in P$. CRUSO uses the Winnowing algorithm \cite{schleimer2003Winnowing} to represent source code as fingerprints, whose length is almost the same as the length of input source code. When used for source code matching, the variable-length fingerprints lead to a large number of source code comparisons, resulting in a significant memory usage and execution time.
\end{enumerate}

To improve the performance of CRUSO, we replaced the Winnowing algorithm with the Paragraph vectors algorithm (PVA) \cite{le2014distributed}, which uses a fixed-length vector representation for source code. We develop a reference dataset that stores the vector representations for code fragments present in SO posts generated using PVA, and the cosine similarity\footnote{\url{http://bit.ly/2ODWoEy}} of all the code fragments from a reference code fragment is chosen randomly. Thus, detecting relevant SO posts to given source code under review becomes a database search query for projecting the SO posts with the similarity score above a specific threshold value. We named the newer PVA-based version of CRUSO as CRUSO-P.

\subsection{Existing techniques for source code representation}
\label{sec:source-code-representation}
The representation of source code plays a significant role while training ML models. A broad categorization of the existing ML approaches based on their representation is as follows:
\begin{enumerate}
    \item \emph{Fingerprint-based approaches}: A typical \emph{code fingerprint} is a compact collection of integers, which summarizes the source code's critical aspects.  The fingerprint-based approaches make use of code fingerprints generated by different algorithms to detect the source code similarity. The algorithms used generally comprise of Winnowing algorithm \cite{schleimer2003Winnowing} and hash-based methods such as MD5 and SHA-1. Winnowing has been used as a source code similarity detection in software activities such as \emph{plagiarism detection} \cite{djuric2013source} and \emph{code review} \cite{Sodhi2019UsingSC}. Similarly, MD5 and SHA have been used to detect different source code clones \cite{akram2018droidcc}. We use cosine similarity measure to detect the source code similarity between different code fragments. 
    \item \emph{Abstract Syntax Tree (AST)-based approaches}: ASTs capture the syntactical details of programming constructs' used in source code. An AST of source code represents a hierarchical structure (tree) comprising of the programming constructs used in the source code in the order of their usage. However, the usage of AST differs in various research works. For instance, authors in  \cite{wang2016automatically} use a linear collection of programming constructs present in the source code's AST for training a Deep Belief Network to perform defectiveness estimation. In contrast, the authors in \cite{chilowicz2009syntax}  use the AST fingerprints to detect similar source code existences using exact matches' clustering.
    
    \item \emph{Software metrics-based approaches}: \label{point:procon-metrics} 
    Software metrics such as Chidamber and Kemerer's (CK's) OO metrics \cite{hitz1996chidamber} and McCabe’s cyclomatic complexity \cite{mccabe1976complexity} have been used to extract source code specific information from various Open Source Software (OSS). Such software-specific information is used to develop large datasets (such as PROMISE repository \cite{menzies2012promise}), which are used to train ML models to perform defectiveness estimation. Programming Construct (PROCON) metrics proposed by authors in  \cite{kapur2020defect} capture the usage patterns of programming constructs occurring in source code. The specific programming constructs are fetched by using the AST generated by parsing the source code. Authors \cite{kapur2020defect} show that the defectiveness estimation performed using PROCON metrics and the state-of-the-art ML technique produce effective results and outperform the existing methods \cite{wang2016automatically}, \cite{hitz1996chidamber}, and \cite{menzies2012promise}.
\end{enumerate}

\textbf{Need and opportunity for newer methods:} 
The fingerprint approaches have the limitations of high processing cost and storage requirement. In contrast, the AST-based techniques generally train the ML models using binary classifiers, such as SVM. The binary classifiers, however, focus on classifying the input source files as \{defective, unpredictable\} \cite{kapur2020defect}, but do not provide any information for the source files \textit{unlikely to be defective}. In our previous work \cite{Sodhi2019UsingSC}, we provide a method for estimating the source files \emph{unlikely to be defective}, but the method used is slower and inefficient. Thus, a faster and efficient method is desirable. Further, to the best of our knowledge, there do not exist efficient algorithms for assisting code reviews. On the other hand, in our current work, we compute fixed-size vector representations of source code. Computing the similarity of two vectors is more efficient and speedy in comparison to that in the case of fingerprints.

\begin{figure}
\includegraphics[width=\columnwidth]{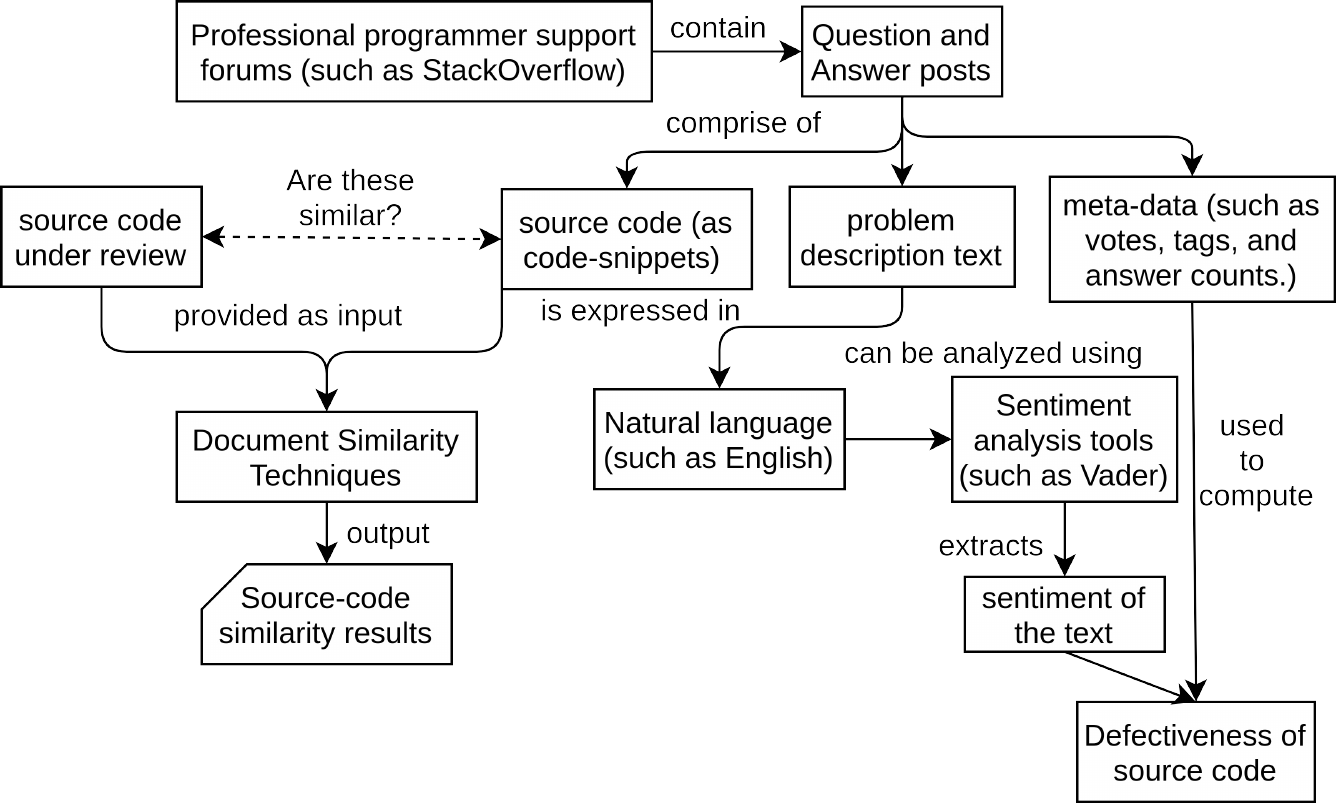}
\caption{Basic idea of our approach}
\label{fig:basic-idea}
\end{figure}

\subsection{Basic tenets behind our system}
\label{sec:basic-tenets-of-our-system}
The basic idea of our approach is stated as follows:
\begin{enumerate}
    \item Professional programmer support forums such as StackOverflow, contain useful information about programmers' problems when developing software. The information includes code fragments, the associated problem, and solution discussions in a natural language.
    
    \item If a \textit{significant} part of given source code under review $c'$ is found \textit{sufficiently similar} to the code fragments $c_p$ present in a SO post $p$, we can infer the \textit{defectiveness} of the $c'$ by analyzing the information available in $p$. We estimate a post's defectiveness by considering its natural language text, various metadata such as up-votes, and the post type.
\end{enumerate}

\subsection{Leveraging crowd-knowledge to identify problems in source code}

\begin{figure}
\centering
\includegraphics[width=\columnwidth]{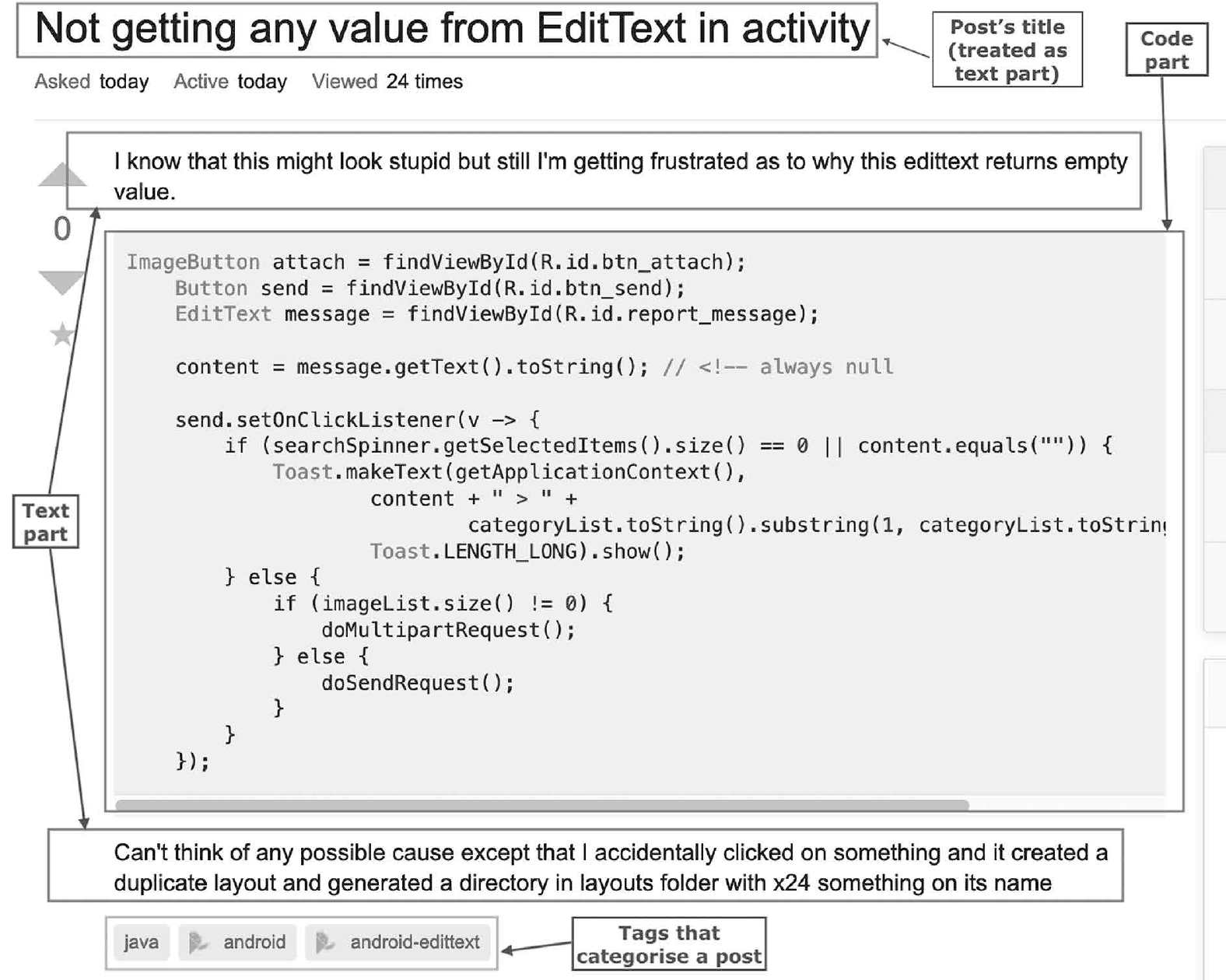}
\caption{Example of a post on StackOverflow}
\label{fig:so-post-example}
\end{figure}

Professional programmer support venues such as 
SO, provide a platform for programmers to discuss various problems related to software development. Figure \ref{fig:so-post-example} shows an example SO post. In this example, the programmer has posted a fragment of source code in which he faces some problem. The description of the problem is available in the post's narrative, which is written in English. Further, the posts are categorized with tags fields, which provide information about the technologies, or the platforms related to the post's code.

SO offers a rich and large corpus of such natural language discussions and the source code fragments discussed in software development-related issues. Works such as \cite{Sodhi2019UsingSC, ponzanelli2013seahawk, ponzanelli2014mining} have shown that it is possible to exploit the crowd-knowledge available at SO for developing tools that address various software development tasks.  Thus, our approach makes use of the rich volume of SO content to identify potential problems in a given source file, which is under review.

Association between software development and crowdsourced knowledge has been studied and confirmed by authors in \cite{stackoverflow-github}, where they studied data from GitHub (an accessible repository of OSS) and StackOverflow. The type of questions that are asked and get answered or remain unanswered on StackOverflow has been explored by \cite{stackoverflow-topics, stackoverflow-empirical-study}. There are mainly two types of posts on StackOverflow:
\begin{enumerate}
    \item Questions posted by programmers soliciting help and solutions for a programming or design problem they face with code or API. We refer to such posts as the \emph{question posts}.
    \item Replies posted by other experts for the above type of posts. We refer to such posts as the \emph{answer posts}.
\end{enumerate}
The analysis of SO data shows\footnote{Our query is available at https://bit.ly/2JSSMez} that the count of Type-2 posts is more than 1.5 times\footnote{On StackOverflow (SO), there are more than 19m questions and 29m answers as of April 2020. See https://data.stackexchange.com/.} the count of Type-1 posts. Further, we find\footnote{Our query for finding this number is available here: https://bit.ly/3c4P79y.} that more than 16\% of Type-1 posts contain source code fragment(s). In contrast, more than 12\% of Type-II posts\footnote{Our query for finding this percentage is available at: \url{https://data.stackexchange.com/meta.stackoverflow/query/edit/1223740}} contain source code fragment(s). These SO posts typically describes some problems involving the source code fragment(s) included in the post. 

Given the above, it can be argued that i) A code fragment accompanying a SO question is quite likely to be involved in a defect \cite{stackoverflow-topics, stackoverflow-empirical-study}, and ii) The code accompanying accepted or high scoring SO answers to such a question is quite likely to be free from the associated question post.

\textbf{Challenges and opportunities:}
Though the SO provides a trove of information about the issues faced by professional programmers during software development, exploiting that information to build code review assistant tools poses several challenges. Major ones include:
\begin{enumerate}
    \item Accurate identification of the SO posts that contain source code fragments matching the input source file.
    
    \item Efficient retrieval of the matched or relevant SO posts.
    
    \item Accurately determining the defectiveness of SO posts.
\end{enumerate}

\begin{figure}[ht]
\begin{subfigure}{\textwidth}
  \includegraphics[width=0.5\columnwidth]{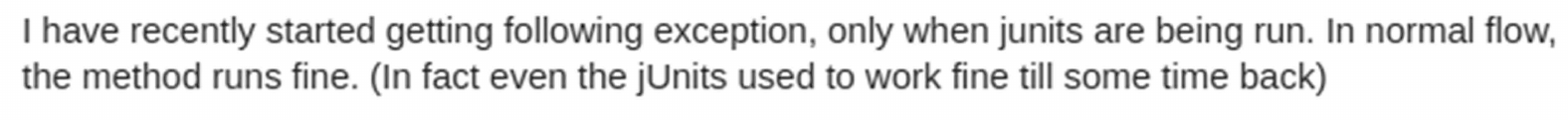}
  \caption{narrative from a SO post}
  \label{fig:example2}
\end{subfigure}

\begin{subfigure}{\textwidth}
  \includegraphics[width=0.5\columnwidth]{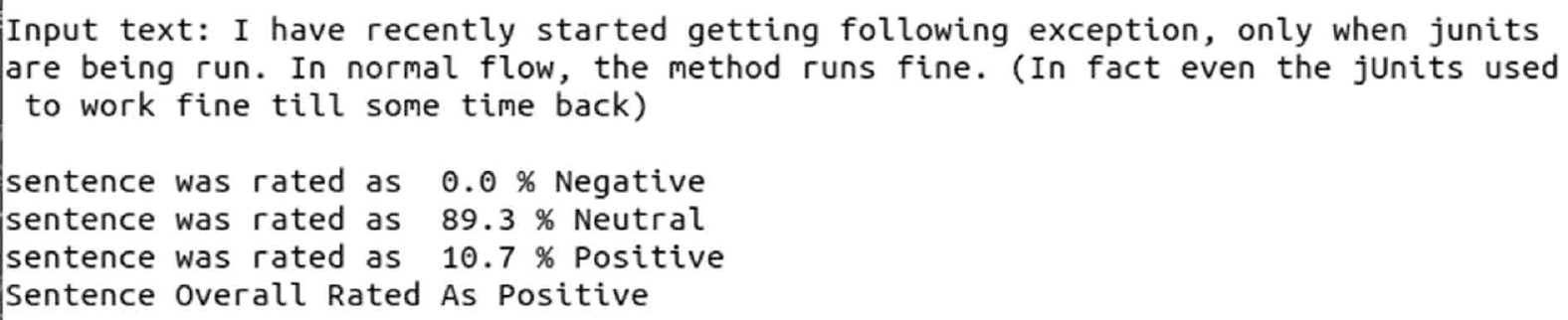}
  \caption{Results from VADER}
  \label{fig:VADER-results}
\end{subfigure}%
\caption{An example of misclassification by the existing Sentiment Analysis tools}
\label{fig:misclassification}
\end{figure}

\textbf{Addressing the above challenges:}
While we note the above challenges, there exist techniques that can be exploited to address them. For an input source code to be reviewed ($c'$), we address the existing challenges:

\begin{enumerate}
    \item \emph{Identifying SO posts containing similar source code to $c'$}: The Paragraph Vector algorithm (PVA) \cite{le2014distributed} has delivered state-of-the-art results \cite{par-vec-nips15} in many Natural Language Processing (NLP) tasks that require a vector representation of text. One of this work's goals is to evaluate the effectiveness of the well-known PVA in computing an accurate representation of source code. Having such a representation of source code is useful for performing efficient and accurate source code comparisons. 
    \item \emph{Efficient retrieval of the matched SO posts ($P'$):} Most of the existing source code comparison methods, such as \cite{Sodhi2019UsingSC,zimmermann2009predicting}, have high processing time and storage requirements. Thus, there is a need to provide a code review solution that accelerates the process and has a lower storage requirement.  
    
    \item \emph{Accurately determining the defectiveness of SO posts:} To determine the defectiveness of a SO post, one can analyze the post's narrative's sentiment. Tools such as CoreNLP   \cite{manning2014stanford} and Valence Aware Dictionary and sEntiment Reasoner (VADER) \cite{hutto2014vader} can be used to infer the \emph{sentiment} of a given input text, but may not prove to be effective when used in the context of some domain-specific narrative. For instance, consider the SO post narrative\footnote{SO post considered as an example: \url{https://bit.ly/2UvZXyg}} shown in Figure \ref{fig:example2}. The sentiment analysis tools, such as VADER,  classify the post as positive with the sentiment score of 10.4\%, which is thus a misclassification. The results obtained are shown in Figure \ref{fig:VADER-results}. Therefore, it is inadequate to rely on a SO post's narrative text to compute the code's defectiveness embedded in it solely.
\end{enumerate}

\section{Proposed approach}
\label{sec:approach}Our system's primary goal can be stated as follows: \textit{Given a source file $f$ written in a programming language $\lambda$, determine if $f$ is likely to have semantic issues.} The central idea behind our approach to addressing the above goal is to \textit{look for any existing source code, which is \underline{sufficiently similar} to the source code present in $f$, and is known to \underline{have a semantic issue}.}

Thus, two tasks become crucial for our approach a) determining the similarity of two source code samples, and b) establishing that the given source code is Likely-to-be-defective. SO is a widely used channel for professional programmer support. It offers a rich corpus of question and answers reply with relevant source code fragments. 

\begin{table}[htbp]\caption{Table of Notation}
\centering 
    \begin{tabular}{r c p{7cm} }
    \toprule
    $L$ & $\triangleq$ & The set of programming languages \{C, C\#, Java, JavaScript, Python\}. We consider the source files written in any one of these. \\
    
    $G$ & $\triangleq$ & Set of considered GitHub repositories, containing source files written in $L$.\\
    
    $S$ & $\triangleq$ &Set of source files in $G$ that are written in $L$.\\
    
    $M$ & $\triangleq$ & The set of PVA models trained using $S$.\\

    $T$ & $\triangleq$ & Test-bed used for testing the performance of $M$. \\
    
    $D$ & $\triangleq$ & Database containing code, text, metadata and other computed items for SO posts.\\

    $R$ & $\triangleq$ & The set of reference vectors chosen for programming languages $\lambda \in L$.\\

     $I$ & $\triangleq$ & Set of metadata items of the SO posts.\\

    $P$ & $\triangleq$ & Set of PVA parameter variation scenarios.\\
    
    $p_{\lambda}$ & $\triangleq$ & An SO post containing $k$ code fragments written in a language $\lambda \in L$. Here, $k > 0$.\\

    $c$ & $\triangleq$ & A code fragment present in  $p_\lambda$.\\

    $v$ & $\triangleq$ & PVA computed vector representation of $c$.\\
    
    $\alpha$ & $\triangleq$ & Cosine similarity between two PVA vectors $v$ and $v'$.\\
    
   $\hat{\alpha}$ & $\triangleq$ & The threshold of cosine similarity between two PVA vectors to categorize them as similar.\\
    
    $\mu$ & $\triangleq$ & The threshold for score metadata field value of various SO posts.\\

    $\psi$ & $\triangleq$ & No. of training samples used for training a PVA model. \\
    
    $\gamma$ & $\triangleq$ & PVA vector size. \\
    
    $\beta$ & $\triangleq$ & No. of training iterations or epochs used for training a PVA model.\\
     $\chi$ & $\triangleq$ & Sentiment score of different sentiment values provided as output by VADER.\\
    
    \bottomrule
    \end{tabular}
\label{tab:CRUSO-TableOfNotation}
\end{table}

\begin{figure*}
\centering
\includegraphics[width=1.4\columnwidth]{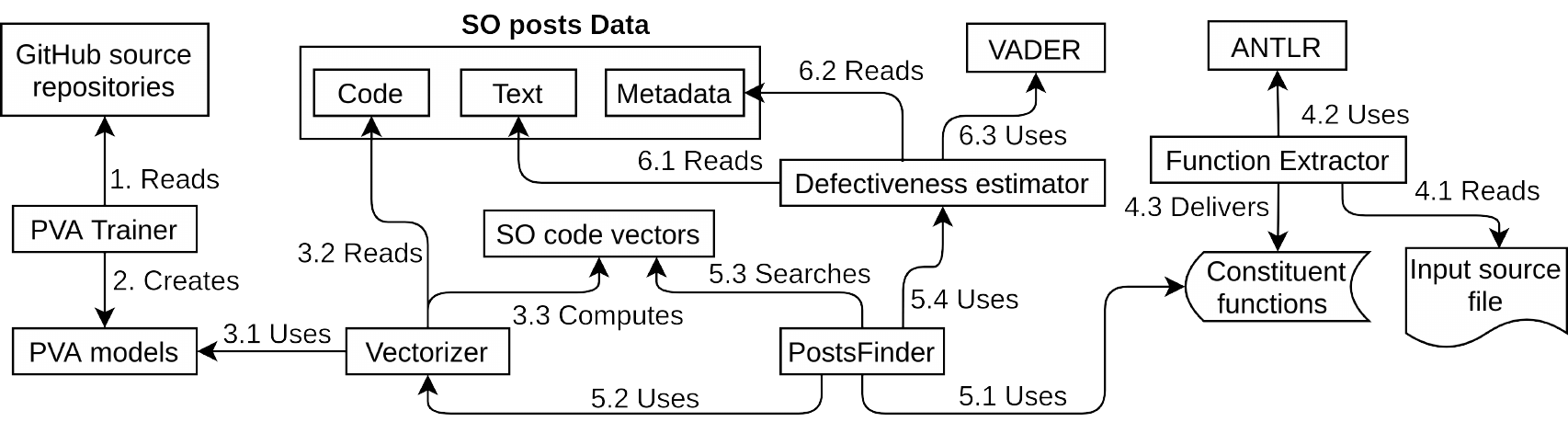}
\caption{Architecture of the proposed system}
\label{fig:system-arch}
\end{figure*}

Table-\ref{tab:CRUSO-TableOfNotation} shows the notation used for various terms in this paper.

\subsection{Steps in our approach}
\label{sec:CRUSO-steps}
Figure \ref{fig:system-arch} shows the architecture of the proposed system that implements our approach, and the critical steps in our approach are listed as follows.
Along with each step, we highlight the relevant design decisions that were addressed when implementing those steps.

\begin{enumerate}
    \item \textbf{Preparing SO posts and code vectors}
    \begin{enumerate}
    \item Download a data dump of SO posts.
    \item Extract the code, text, and metadata parts from each SO post $p_\lambda$ and store in a database $D$.
    
    \textit{Design decision: How to decide whether a SO post and its content are relevant and useful?}
    
    \item Download source files from GitHub repositories, such that they are written in a programming language $\lambda \in L$. 
    
    \textit{Design decision: Why only GitHub? How do we select a source file? Why only these programming languages?}
    
    \item For each language $\lambda \in L$, train PVA models using the samples from GitHub source files. 
    
    \textit{Design decision: Why use GitHub source files for training? How to decide which files to choose from them? Why use PVA and how to choose the values of its tuning parameters?} 
    
    \item For each SO posts' code fragment $c_i$ available in $D$, compute its vector representation $v_i$ using a suitable language-specific PVA model trained above. The vector is stored along with the corresponding code fragment in $D$ itself.
    
    \textit{Design decision: Why use vector representations of source code?}
    
\end{enumerate}

\item \textbf{Determining the defectiveness of source code under review ($c'$)}
\begin{enumerate}
    \item Compute the vector representation $v'$ for the input source code $c'$ after suitably preprocessing it.
    
    \item Find all vectors $v \in D$ such that the similarity ($\alpha$) between $v$ and $v'$  is above a similarity threshold ($\hat{\alpha}$). 
    
    \textit{Design decision: What should be the value of $\hat{\alpha}$? On what factors does it depend?}
    
    \item For each $v \in D$, compute the defectiveness value, $\delta$, using the narrative and metadata of the SO post $p_\lambda$ of $v$.
    
    \textit{Design decision: How to compute $\delta$?}
    
    \item The defectiveness value $\delta'$ for the input source code $c'$ is computed by considering all $\delta$ of $\forall v \in D$.
    
    \textit{Design decision: How to compute $\delta'$?}
    
\end{enumerate}

\end{enumerate}

 We discuss the crucial design decisions faced in our approach in the next subsection.

\subsection{Design considerations in our approach}
\label{sec:DesignConsiderations}
In the following subsections, we describe the details of the steps involved in developing our software artifacts and the rationale for design decisions addressed at each step.

\subsubsection{Selection of the SO posts and  the programming languages}
The SO posts have mainly three types of content: i) questions, ii) answers, and iii) comments and metadata of the post. Further, two SO questions (or answers) may not have the same level of detail. For instance, a question post may have very little or no source code present in it. Alternatively, a post may have multiple code fragments written in different programming languages. Similar issues are present for other types of SO posts. Thus, it becomes essential to decide \textit{if a SO post and its content are relevant and should be considered or not.} 

\textbf{Selecting the suitable SO posts:}
To address the above question, we adopt the following criteria for selecting SO posts for our use: 
\begin{enumerate}
    \item \emph{Size constraint:} The size of the code fragment(s) present in the post should be greater than a certain threshold ($>100$ characters without white spaces). We assume that any source code performing a logical function is of size above this threshold.
\item \emph{Tag constraint:} The post should be tagged (i.e. categorised) with at least one of the programming languages that we consider. 
\end{enumerate}

\textbf{Selecting suitable programming languages:} 
Following factors were considered when selecting the programming languages: 

\begin{enumerate}
    \item We should cover multiple programming paradigms, such as object-oriented, procedural, and scripting.
    \item The languages should have a significant active deployment in the field.
\end{enumerate}

After surveying the existing literature \cite{so-survey-2019} and studying the online trends of developer-usage\footnote{Sources of stats: https://githut.info/}, we arrived at the set $L = $ \{C, C\#, Java, JavaScript, and Python\} of actively used programming languages.

\subsubsection{The rationale for choosing PVA}
A crucial design decision that requires explanation is the use of PVA in our approach. The following are the main reasons for our choice of PVA: i) It allows us to compute vectors of the same length that accurately represents the source code samples. Keeping the length of such vectors the same for every source code sample is critical for implementing an efficient and fast system. ii) Our experiments show that the PVA works equally well for all the programming languages that we considered. iii) Recent works such as \cite{Code2VecAlon2019}, a close variant of PVA, have proven that it is possible to compute accurate vector representations of source code and that such vectors can be very useful in computing semantic similarity between two source code samples.
Thus, we chose the PVA to compute vector representations of source code samples in our approach.

\subsubsection{The rationale for using source code samples from GitHub and SO posts data}
\label{sec:selecting-GitHub}
For training the PVA models, we used source code samples taken from various GitHub repositories. We used source code from GitHub repositories due to the following reason: 
To train the PVA models with realistic source code samples that we expect to encounter in real-world usage, we use source code samples from GitHub repositories. The source code samples taken from GitHub are syntactically complete units (e.g., a complete Java class instead of just a method definition).

A related question is \emph{which source files to choose from GitHub?}. We randomly selected repositories which met the following criteria:
    \begin{enumerate}
        \item The repository had the source files written in a programming languages $\lambda \in L$ (see Table-\ref{tab:CRUSO-TableOfNotation}).
        \item The repository had earned 100 or more stars.
        \item The repository had more than 1000 source files.
    \end{enumerate}

Based on the above criteria, we selected about 105 different repositories\footnote{Details can be found in our dataset script available at \url{http://bit.ly/2KJVWCh}} on GitHub from which the source files were taken for training and testing of the PVA models. Table \ref{table:cloc-stats} presents the detail of GitHub source files selected corresponding to various programming languages to train the PVA models. We use the \textit{cloc} tool \cite{cloc} to compute the count of comments, blank lines, and source lines of code (SLOC) present in the source code. 

We use the SO posts data to create a reference dataset to perform the source code matching during the code review process. Using the PVA models, we obtain the vector representations for code fragments present in various SO posts and store them in a relational database to perform the vector comparisons. We choose the SO posts data for the following reason: The code present in SO posts is a mix of syntactically partial code fragments and full ones. For example, some posts contain the complete Java class definitions, while others may contain only a method definition or a small code block. On the other hand, the input code that we want to check for defectiveness will almost always be a syntactically complete unit of source code, such as a Java class or a Python module. 

\begin{table}
\centering
\caption{Details of the dataset used for training and testing of PVA models}
\label{table:cloc-stats}
\resizebox{\columnwidth}{!}
    {
\begin{tabular}{l r r r l | r r r}
\cmidrule[\heavyrulewidth]{2-8}
&\multicolumn{4}{l|}{\textbf{Training corpus (lines of code measured by \texttt{cloc})}} &
 \multicolumn{3}{c}{\textbf{Testing corpus}}  \\ \hline
\textbf{Language} & \textbf{Files} & \textbf{Blank} & \textbf{Comment} & \textbf{SLOC} 
& \textbf{File pairs} & \textbf{Corpus size} & \textbf{Models tested} \\ \midrule
C & 32099 & 2908784 & 2490163 & 14908295 & 5000 & 30036 & 21 \\
C\# & 8112 & 303416 & 198693 & 2342959 & 5000 & 7076 & 21 \\
Java & 142266 & 437851 & 659172 & 2157881 & 5000 & 127568 & 21 \\
JavaScript & 15737 & 177587 & 226724 & 1259902 & 5000 & 12485 & 21 \\
Python & 6012 & 300109 & 412452 & 1248494 & 5000 & 5378 & 20 \\ \bottomrule
\end{tabular}}
\end{table}

\subsubsection{Choosing PVA parameters and code similarity threshold, $\hat{\alpha}$}
Performance, in terms of accuracy, storage efficiency, and response time, is determined by its input parameters such as $\beta, \gamma$, and $\psi$ (see Table-\ref{tab:CRUSO-TableOfNotation}). Therefore, one of the key challenges in using PVA in our system is determining the minimum threshold value of  $\alpha$, indicating a significant similarity between two source code samples. Further, we would also like to select the optimal values of $\beta, \gamma$, and $\psi$ that can result in such a value of $\hat{\alpha}$ (see Table-\ref{tab:CRUSO-TableOfNotation}).
The details of the experiments performed to determine the optimally tuned values of $\beta, \gamma$, $\psi$ and, $\hat{\alpha}$ for obtaining the best performing PVA models are provided at \url{https://bit.ly/2Ig3crd}. 

\subsubsection{Computing defectiveness, $\delta$}
\label{sec:compute-defectiveness}
The essential parameters considered while computing the defectiveness are:
\begin{enumerate}
    \item \emph{The sentiment of a post's narrative:} This describes the view or opinion of the problem and is computed using the VADER sentiment analysis tool. A post's sentiment can be \emph{positive}, \emph{negative}, or \emph{neutral}, depending on the problem's narration. A SO post consisting of a \emph{negative} narration is most likely to describe a problem, thus comprising a defective code. Similarly, a SO post with a \emph{positive} narrative is most likely to propose a solution code for a programming problem.  
    
    \item \emph{The score value of the post:} This is an integer value, available with every SO post, describing the approval or disapproval of the post by various viewers. An approval increments the score value, while the disapproval decrements it. A post with a high score value reflects a considerable confidence value in the post's content. For instance, the source code present in a high score answer post is likely to be free from defects.
    
    \item \emph{The type of SO post:} A SO post can be classified as a question post or an answer post. A \emph{question} post generally projects a programming problem containing a \emph{source code defect}. On the contrary, an \emph{answer} post mostly provides a solution source code, which is \emph{unlikely-to-be-defective}.  
\end{enumerate}
To compute the defectiveness of a SO posts' code snippet $c_p$, we combine the results obtained based on $p$'s metadata (viz., score ($\mu_{p}$) and  its post-type) and the sentiment information of the narrative in $p$. The complete procedure for computing the defectiveness ($\delta$) of a code snippet $c$ present in a SO post $p$ is listed in Algorithm \ref{alg:computing-defectiveness}.  To obtain the thresholds of SO post's score values, we compute the statistical measures, viz., maximum ($max$), minimum ($min$), average ($avg$), and standard deviation ($stddev$) of the respective score ($\mu$) values of different types of SO posts. 
To find the source code's programming language present in a SO post, we use the \emph{tag} metadata field. We choose the $avg(\mu)$ values under each of the language category and the post types as the respective thresholds. We represent the threshold values for question and answer posts as $avg(\mu_{q})$ and $avg(\mu_{a})$. By observing the threshold values, we select the $\langle avg(\mu_{q}), avg(\mu_{a} \rangle$ as $\langle 1, 1.9 \rangle$.   

The defectiveness score computed based on the narrative sentiment of a SO post $p$ is represented as $\delta_p^{narrative}$. The $\delta_p^{narrative}$ values assigned for the sentiment outcomes \{negative, positive, neutral\} are \{-1,1,300\}, respectively. We label a code snippet $c_p$ as \textit{unpredictable} if it has a \emph{neutral} narrative sentiment of post $p$, and $\mu_{p}$ is below the respective post-type thresholds ($\mu_{q}$ or $\mu_{a}$). Also, the $\delta$ values corresponding to various defectiveness labels \{Likely-to-be-defective, Unlikely-to-be-defective, Unpredictable\} are \{-1,1,300\}, respectively.    
To avoid the false negatives, we compute the $min$ defectiveness score on Step \ref{step:taking-min} of Algorithm \ref{alg:computing-defectiveness}. We define the false negative as a case when a defective source code gets labeled as non-defective. A high value of $\delta_p^{narrative}$ for \emph{neutral} sentiment value is chosen to consider the defectiveness inferred from the score metadata field of $p$.

\begin{algorithm}[ht]
\caption{Steps for computing $\delta$ of a code snippet $c$ present in a SO post $p$}
  \label{alg:computing-defectiveness}
\begin{algorithmic}[1]
\REQUIRE 
 $I = $ Set of metadata items associated with the StackOverflow post $p$. \\
  $t = $ The narrative text present in $p$. \\
  $\mu_q,\ \mu_a  = $ The threshold score values for question and answer posts of SO respectively.\\
\ENSURE   $\delta_p =$ The defectiveness score for code snippet $c$ present in $p$.

\STATE{$\delta_p =0$}
\IF{I(p.PostType) = question and I(p.Score) $> \mu_q$}
    \STATE{$\delta_p =$ -1}
\ELSIF{I(p.PostType) = answer}
    \IF{I(p.Score) $> \mu_a$}
        \STATE{$\delta_p =$ 1}
    \ELSE
        \STATE{$\delta_p =$ -1}
    \ENDIF
     \STATE{$\delta_{p}^{narrative} =$ computeNarrativeSentiment($t$)}\COMMENT{Using VADER}\\
     \STATE{$\delta_p \label{step:taking-min} =$ min($\delta_p, \delta_{p}^{narrative}$)}\\
     \STATE{saveToDatabase($p.id, \delta_p$)}\\
\ENDIF
\end{algorithmic}
\end{algorithm}

\subsection{Implementation details}
\label{sec:implementation}

    
    

We developed our software artifacts using the programming languages - Java and Python. These languages were selected because of  the available expertise.      

\subsubsection{Developing the SO posts database -- SOpostsDB} 
\label{sec:database-impl}

SOpostsDB forms the knowledge base of our code review assistant system. The significant steps involved in developing the database are as follows:

\begin{enumerate}
    \item \emph{Data Collection}: We downloaded a dump of SO posts \cite{stack-exchange-dump} between July 2008 and March 2018 containing more than 5 million posts\footnote{\url{https://archive.org/download/stackexchange}}. We extracted various sub-components (viz., code, text, and metadata) of these SO posts and store them in the database as the \emph{SOpostsData} table. We also downloaded 105 OSS repositories from GitHub, containing different source files written in various programming languages.
    
    For developing the SOpostsDB, we considered the SO posts containing source code written only in five programming languages, viz., C, C\#, Java, JavaScript, and Python. We collected about 188200 source files in total from GitHub written in these programming languages. 
    We used the data collected from GitHub repositories to train our PVA models, while the source code extracted from SO posts to perform the source code matching with given source code. The link to access the training data and the SO posts details is provided at \url{https://bit.ly/39KiA7l}.  
    
    \begin{figure}
\centering
\includegraphics[width=0.6\columnwidth]{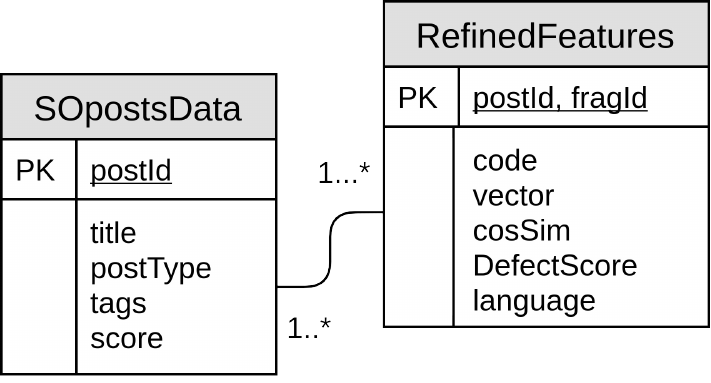}
\caption{Relational Schema of SOpostsDB}
\label{fig:relational-schema}
\end{figure}

Figure \ref{fig:relational-schema} shows the relational schema of our database. A SO post $p$ may consist of multiple code fragments $c_i$ surrounded by different narratives $t_i$. We consider each of such $c_i$ and the $t_i$ preceding $p$ as a single fragment and represent all such fragments with distinct \texttt{fragId}(s). \textit{In other words, we consider the \underline{code fragments} the \underline{basic unit of source code comparison} and thus obtain the PVA vectors corresponding to them}. We consider the SOpostsData and CodeVectors as two separate tables because \textit{a)} the SOpostsData comprises of the initial preliminary information required to obtain the information present in the CodeVectors table, and \textit{b)} The SOPostsData majorly comprises of the metadata fields of a post which are common to all the code fragments present in a post. 

\item \emph{Storing the vector representations of the code fragments present in SO posts}: The code fragments extracted from SO posts are preprocessed and stored in the database. To expedite the process of source code comparison, we perform the following:

\begin{enumerate}
    \item \label{step:obtain-vector-rep} Using language-specific PVA models, we obtain the vector representations of the extracted code fragments and store them. We use the python implementation of the \texttt{gensim}  library \cite{gensim} to implement PVA, preprocess the source code, and obtain the vector representations corresponding to them.    The vector representations obtained for the code fragments present in various SO posts are stored in the database' \emph{CodeVectors} table.
    \item For each of the considered programming languages $\lambda \in L$, we select a PVA vector and refer to it as a \emph{reference vector}. To select the reference vectors, we order the vector representations $v$ present in the database using the \texttt{postId} and the \texttt{fragId} of the records and select the first entry for each of the languages as the reference vectors $v_r$.  
    \item We store the cosine similarity\footnote{\url{https://scikit-learn.org/stable/modules/generated/sklearn.metrics.pairwise.cosine_similarity.html}} (\texttt{cosSim}) of the PVA vectors (obtained in step \ref{step:obtain-vector-rep}) with the respective reference vectors $v_r$. We use the \texttt{sklearn} library's python implementation \cite{scikit-learn} to compute the cosine similarity measure values between two vector representations. 
\end{enumerate}

\textbf{The idea behind storing the cosine similarity with the reference vectors:} Consider the following scenario:
\begin{enumerate}
    \item If a PVA vector $v$ of a code fragment $c$ present in the database, has a high cosine similarity score $\alpha$ with the reference vector $v_r$, and
    \item The PVA vector $v'$ of an input source code $c'$ also represents a high cosine similarity score with $v_r$, then
    \item The code fragments $c$ and $c'$ are likely to have a high degree of code similarity between them.
\end{enumerate}

The queries to the \textit{CodeVectors} table for finding vectors that are \emph{cosine similar} to a given vector are expensive if done on the vectors. To make such queries efficient, we pre-compute each SO code vector's cosine similarity with a language-specific reference vector and store it in the database. The cosine similarity value is a scalar quantity, and we can create an index on this relational attribute to speed up the database queries.

A smaller version of our database, comprising the \textit{CodeVectors} table and a subset of metadata fields (the SO post's title field) in \textit{SOPostsData}, used by our tool, is shared \url{https://bit.ly/2xs8CtV}.
\end{enumerate}

\subsubsection{Developing the Code review system using SO and PVA -- CRUSO-P}
\label{sec:cruso-working}
CRUSO-P is an automated solution for assisting the code review activity. For providing this assistance, CRUSO-P utilizes the knowledge accumulated in SOpostsDB and code-similarity models trained using PVA. For a given input source code $c'$, CRUSO-P labels $c'$ as \{Likely-to-be-defective, UnLikely-to-be-defective, Unpredictable\}. The major sub-modules driving CRUSO-P and their implementation details are described next. 

\begin{enumerate}
    \item \emph{PVA Trainer}: This module is responsible for developing the PVA models. The best performing model corresponding to  each of the programming languages $\lambda \in L$ is found by running several experiments. 
    To find the optimal values of the input parameters $\beta, \gamma,$ and $\psi$, the PVA trainer learns different models built using various parameter combinations. The \texttt{doc2vec} function of the \texttt{gensim} library is used to learn various PVA models. Several experiments were performed to obtain the optimal values of $\beta, \gamma,$ and $\psi$, and to determine the similarity threshold values $\hat{\alpha}$ for source code written in different programming languages. We deem two source files as highly similar or identical when the PVA similarity score ($\alpha$) is higher than a threshold value, $\hat{\alpha}$. For each of the considered programming languages $\lambda \in L$, the best performing models are found using various evaluation metrics. The details of the experiments performed to obtain the best performing PVA models are discussed at \url{https://bit.ly/2Ig3crd}.
    
    \item \emph{Vectorizer:} This module is used to obtain the vector representations of code fragments present in the SO posts. 
    The fixed-length vector representation of source code improves storage utilization and makes the search and retrieval process efficient. The best performing PVA models corresponding to each of the programming languages $\lambda \in L$ are used to obtain the vector representations. The \texttt{infer\_vector} function of the \texttt{gensim} library is used to obtain the vector representations.

    \item \emph{Posts Finder:} This is the main module that interacts with the front-end tool to provide the defectiveness estimates for an input source code $c'$. The SOPostsDB used for performing the source code matching comprises the vector representations corresponding to the code fragments present in the SO posts. The code fragments are generally of the form of code blocks or function bodies. 
    For the reviewed input source files, \textit{we consider a function-definition as the basic unit of source code} \textit{comparison.}
    
    For the input source files that are reviewed, \textit{we consider a \underline{function-definition} as the \underline{basic unit of source code comparison}}. Posts Finder uses the \emph{Function extractor} to obtain the constituent functions ($W$) present in an input source code $c'$. For each of the obtained constituent functions $\omega \in W$, Posts Finder performs the following:
    \begin{enumerate}
        \item It uses the \emph{vectorizer} module to obtain the PVA vector corresponding to $\omega$, say $v_\omega$.
        \item It obtains the cosine similarity score $\alpha$ between $v_\omega$ and the language-specific reference vectors $v_r$. 
        \item It fetches the top K matching code fragments from the database. The matching code fragments are obtained by fetching the top-K PVA vectors $V$ having $\alpha'$ closest to $\alpha$. We set K = 5 for our tool.
        \item Uses the \emph{defectiveness estimator}  to obtain the matching code fragments' defectiveness estimates and thus of $\omega$.
         
    \end{enumerate}
    
    The final estimate on the defectiveness of $f$ is taken by performing a majority vote on the constituent functions' defectiveness estimates. \emph{We compute the majority vote defectiveness estimate by computing the statistical} \texttt{mode} \emph{of the defectiveness values obtained for the constituent functions $W$ of $f$}. The complete procedure followed in computing the defectiveness of the input source code $c'$ is listed in Algorithm \ref{alg:CRUSO-testing}.
    
     \item The \emph{function extractor:} This module's goal is to extract the constituent function definitions present in an input source code. The \emph{Function extractor} parses the input source code using \emph{ANTLR}\footnote{https://www.antlr.org/} and builds a custom \textit{Listener} by modifying the function or method call event definitions. We use the Java programming language to build this module and transform it into a JAR executable using the Apache Maven Shade plugin\footnote{http://maven.apache.org/plugins/maven-shade-plugin/examples/executable-jar.html}. This component works as a back-end module in our tool.

     \item \emph{Defectiveness estimator}: This module is used to obtain the defectiveness estimates of the code fragments present in SO posts. For an input code fragment $c$ of a SO post $p$, the \emph{defectiveness estimator} reads the narration and the metadata fields. It uses the VADER tool to compute the narrative sentiment, which can be \{positive (pos), negative (neg), or neutral (neu)\}. For an input text $t$, VADER returns a sentiment score ($\chi$) associated with these sentiment values. The decision function ($\xi$) used to compute the final sentiment value for $t$ is as follows:
    \[{\xi}[t] = 
     \begin{cases}
       \text{positive,} &\quad\text{if $\chi$[pos] $>$ $\chi$[neg] $>= 0.5$}\\
       \text{negative,} &\quad\text{if $\chi$[neg] $>$ $\chi$[pos] $>= 0.5$}\\
       \text{neutral,} &\parbox[t]{0.35\textwidth}{if $\chi$[neu] $>= 0.5$ and $\chi$[pos] $< 0.5$ and $\chi$[neg] $< 0.5$}\\
     \end{cases}
\] The complete procedure to compute the defectiveness of $c$ is listed in Algorithm \ref{alg:computing-defectiveness}.
\end{enumerate}

\begin{algorithm}
\caption{Steps for detecting defectiveness $\delta$ associated with a source-file $f$}
  \label{alg:CRUSO-testing}
\begin{algorithmic}[1]
\REQUIRE 
 $f = $ Source file to check for defectiveness.\\
 $\lambda = $ Programming language in which $f$ is written.\\
 $M = $ Set of PVA models trained for various programming languages ($\forall \lambda \in L$).\\
 $R = $ Set of reference vectors chosen for various programming languages $L$.\\
 $SOPostsDB = $ Database containing the vector representations of code\\ 
 fragments and metadata information of SO posts.\\
\ENSURE   $\delta_f =$ The defectiveness score for source file $f$.

\STATE{$\delta_f = $ 0}\\ 
\STATE{Z = $\phi$}\\
\STATE{$M_\lambda = $ fetchAndLoadModel($M, \lambda$)}\COMMENT{read from the local file system}\\
\STATE{$R_\lambda = $ fetchRefVector($R, \lambda$)}\COMMENT{a query into the SOpostsDB}\\
\STATE{W = parseAndObtainFunctions($f$)}\COMMENT{using Function Extractor}\\
\FORALL{code fragment $\omega \in W$}
    \STATE{$v_\omega = $ obtainVectorRep($\omega, M_\lambda$)}\COMMENT{using Vectorizer}\\
    \STATE{$\alpha_\omega =$ obtainCosSim($v_\omega, R_\lambda$)}\\
    \STATE{$ C = $ fetchTopMatchCodeFrags($\alpha_\omega$, SOPostsDB)}\\
    \COMMENT{Every $c \in C$ has $\langle postId, fragId \rangle$}\\
    \FORALL{code fragment $c \in C$}
        \STATE{$\delta_c = $ obtainDefectivenessEstimate(c.postId, SOPostsDB)}\COMMENT{using Algorithm \ref{alg:computing-defectiveness}}\\
        \STATE{Z = Z $\cup$ $\langle \delta_c \rangle$}\\ 
        \STATE{$\delta_f = $ computeMajorityVoteDecision(Z)}\COMMENT{by computing statistical Mode(Z)}\\
    \ENDFOR
\ENDFOR
\end{algorithmic}
\end{algorithm}

\subsubsection{Testing the defectiveness of source code using the CRUSO-P tool}
\label{sec:testing-source code-using-CRUSO}

CRUSO-P provides a file-uploading interface to the end-user to submit the file to be reviewed. On submitting the source-file f to be reviewed, CRUSO-P outputs the defectiveness decision about $f$ and provides the top matching SO posts results.  The complete testing procedure used by CRUSO-P to detect the defectiveness of $f$ is listed in Algorithm \ref{alg:CRUSO-testing}. CRUSO-P uses the PVA models, and the vector database of SO posts provided as input in this testing procedure. The database also contains the necessary metadata information, such as the type of posts, score of various SO posts, and the defectiveness of various SO posts (computed using Algorithm \ref{alg:computing-defectiveness}).

\begin{figure}
\centering
\includegraphics[width=1.1\columnwidth]{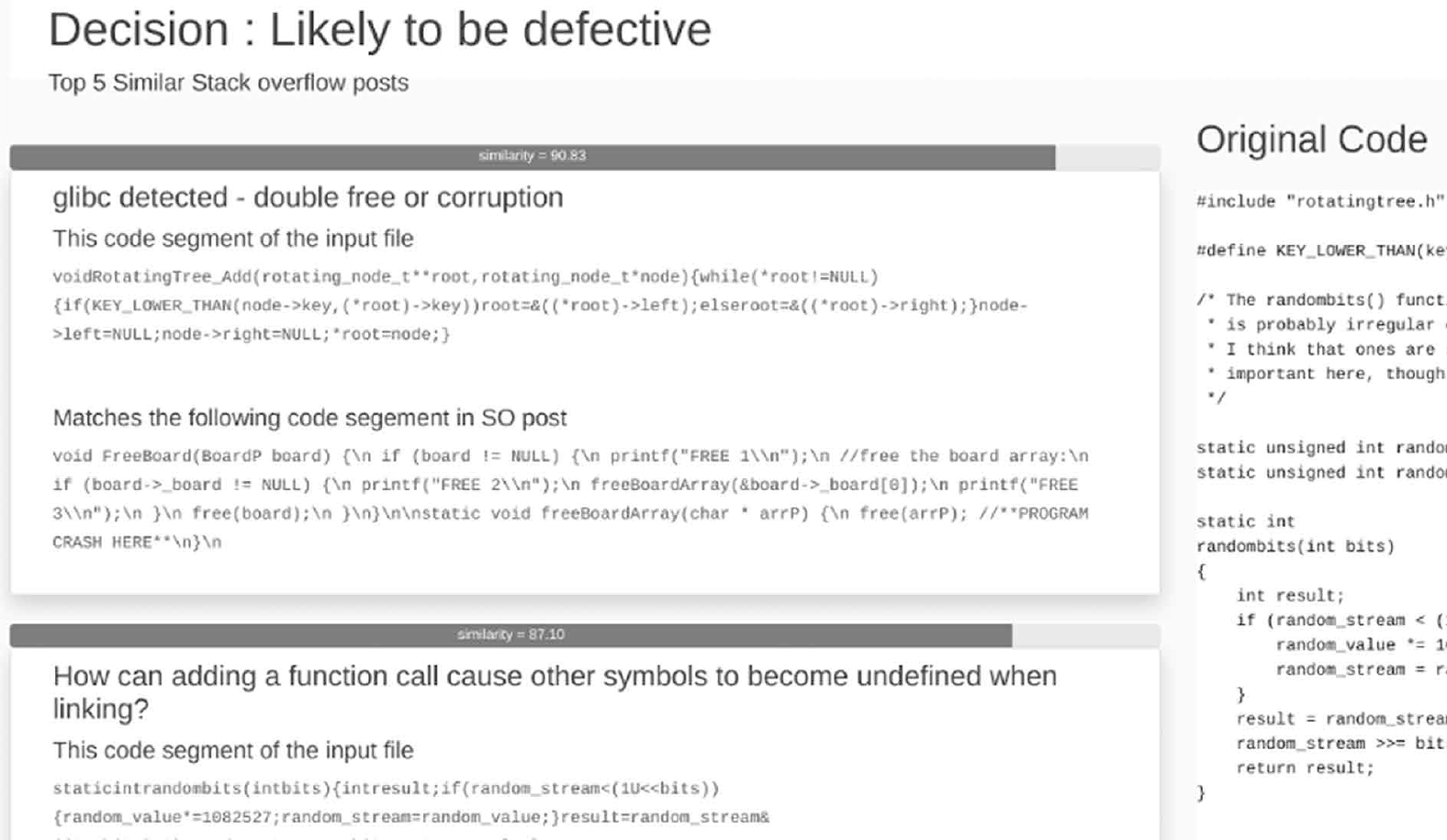}
\caption{Partial view of CRUSO-P code review results}
\label{fig:cruso-code-review-results}
\end{figure}

The complete testing procedure is listed in Algorithm \ref{alg:CRUSO-testing}. Figure \ref{fig:cruso-code-review-results} shows an example of the usage of our tool. Here, we test the tool with an input source file containing the code fetched from GitHub repository cpython\footnote{\url{https://bit.ly/2RyxYxe}}. The figure shows the defectiveness results, the matching code fragments $C$, the associated similarity score, and defectiveness estimates. From the results shown in the figure, 4/5 matching code fragments depicted the defectiveness estimates as \textit{Likely-to-be-defective}, and thus the input post was marked as \textit{Likely-to-be-defective} as per the majority vote criterion. 

It can be validated from the associated defect report link\footnote{https://bit.ly/2yf3RnO} that the source file contains defects, and validates our tool's results. 
Our tool (CRUSO-P) can be accessed at \url{https://bit.ly/2V80NCT}. For a given input source file $f$, CRUSO-P outputs the top matching SO posts' code fragments with their defectiveness estimates and the similarity scores.

\section{Performance Evaluation and Comparison}
\label{sec:evaluation}
One of this work's critical goals is to \textit{significantly improve the speed, efficiency, and accuracy of our previous work \cite{Sodhi2019UsingSC}.} We have achieved this goal by improving the approach for detecting the relevant SO posts for $f$. Our tool determines the relevant SO posts by comparing the cosine similarity among the vector representations of two source codes. The \textit{vectorizer} module of CRUSO-P is responsible for producing the vector representation for an input source code using the pre-trained PVA models. Thus, the accuracy of this task depends on the performance of PVA models. To obtain the best performing PVA models among all the considered programming languages $L$, we performed various parameter tuning experiments (details provided at \url{https://bit.ly/2Ig3crd}).

CRUSO-P infers the defectiveness of an input source code $f$ by analyzing the defectiveness of the similar code fragments present in SOPostsDB. Thus, the performance of CRUSO-P depends on two essential factors:
\begin{enumerate}
    \item Efficacy in detecting the SO posts containing similar code fragments
    \item Precision in computing the defectiveness of SO posts
\end{enumerate}

Therefore, while evaluating the performance of CRUSO-P, we design our experiments around the above two factors. The salient research questions addressed in our experiments are listed below:
\begin{enumerate}
    \item What is the highest accuracy achieved by CRUSO-P? Is CRUSO-P inclined to any specific programming language? 
    \item How does CRUSO-P perform in comparison to CRUSO?
\end{enumerate}

We used the Python programming language to implement our experiments.



\subsection{Evaluation metrics}
\label{sec:evaluation-metrcs}
 \begin{enumerate}
     \item \textbf{Accuracy} is defined as:
    \begin{align}
    Accuracy = \frac{Total\ number\ of\ correctly\ detected\ matches}{Total\ number\ of\ tested\ record\ pairs}
    \end{align}
   
    \item \textbf{F1\ Score} is the harmonic mean of precision and recall:
       \begin{align}
    F1 score = \frac{2\times Precision \times Recall}{Precision + Recall}\end{align}
    
    where 
    \begin{align}
    Precision = \frac{true\ positive}{true\ positive + false\ positive}\end{align}
    
    \begin{align}Recall = \frac{true\ positive}{true\ positive + false\ negative}\end{align}

 \end{enumerate}
 We used the implementation provided by the sklearn library to compute these evaluation metrics.

\subsection{Evaluating the performance of CRUSO-P}
\label{sec:exp-2}
To perform this experiment, we take a subset of code fragments present in our SOposts database and evaluate the performance of our tool in predicting the defectiveness of these code fragments. Our tool can be considered to be effective if it marks a \emph{defective} code fragment as \emph{defective}. and vice versa.

\textbf{The research question addressed:}
\emph{What is the highest accuracy achieved by CRUSO-P? Is CRUSO-P inclined to any specific programming language?}

\subsubsection{Test-bed setup:} 

For our experiment, we selected 5000 code fragments associated with SO posts in a random manner.

\subsubsection{Procedure}
\label{sec:exp-2-procedure}
The procedure to perform the performance evaluation is as follows:

\begin{enumerate}
    \item Test CRUSO-P with the code fragments present in the SO posts, and record the defect estimates provided by CRUSO-P.
    \item Compute the Accuracy score and the F1 score based on the defect estimates provided in the previous step.
\end{enumerate}

\subsubsection{Results and observations}
\label{sec:exp3-results}

\begin{table}
\centering
\caption{Threshold similarity scores and performance scores of CRUSO-P}
\label{tab:cosine-similarity-SO}
\resizebox{0.5\textwidth}{!}{%
\begin{tabular}{l | c c | c c | c |c| }
\cmidrule[\heavyrulewidth]{2-7}
&\multicolumn{2}{c|}{\textbf{Thresholds}}
&\multicolumn{2}{c|}{\textbf{Performance}} & \multirow{2}{*}{\textbf{\shortstack{Number of\\ SO posts}}} & \multirow{2}{*}{\textbf{\shortstack{Time taken\\ (in seconds)}}}  \\ \cmidrule{2-5}
Language & Avg($\alpha$)& StdDev($\alpha$) & Accuracy & F1 score & & \\
      \midrule
 C		&0.963&	0.0704	&0.992&	0.992	&	5000 & 402	\\
C\#		&0.954&	0.0979	&0.8559&	0.8365	&	5000 & 413	\\
Java		&0.97&	0.0668	&0.993&	0.993	&	5000 & 389	\\
JavaScript		&0.967&	0.0719 &0.8766&	0.8612	& 5000 & 451	\\
Python		&0.9617&	0.0764	&0.991&	0.9909	&  5000 & 368	\\

\bottomrule
\end{tabular}
}
\end{table}

\begin{table*}
\caption{Performance comparison of CRUSO-P and CRUSO}
\label{tab:response-time}
\centering
\resizebox{1.4\columnwidth}{!}{
\begin{tabular}{c|c|c|c|c|c|c|c|c} 
 \toprule[1pt]
 \multirow{2}{*}{\textbf{Tool}}&\multicolumn{5}{c|}{\textbf{Programming Language vs. Response time (in seconds)}}& \multirow{2}{*}{\textbf{\shortstack{Avg. Response\\ time (in seconds)}}}&\multirow{2}{*}{\textbf{\shortstack{Storage\\ (in  MBs)}}}&\multirow{2}{*}{\textbf{Accuracy}}\\
 \cline{2-6}
 &\texttt{C}&\texttt{C\#}&\texttt{Java}&\texttt{JavaScript}&\texttt{Python}&&\\
 \midrule[1pt]
 CRUSO-P & 1.09 & 13.15 & 11.47 & 4.35 & 1.35 & 6.28 & 121.53 & 99.3\%\\
 CRUSO & 284.74 & 291.09 & 289.15 & 281.81 & 292.8 & 287.92 & 14239 & 94\%\\
  \bottomrule
\end{tabular}
}
\end{table*}

Table \ref{tab:cosine-similarity-SO} lists the threshold similarity scores and the evaluation metrics (Accuracy and F1 score) values obtained from the experiment.  
\begin{itemize}
    \item \textit{Observations:} The salient observations from the experiment are:
    \begin{itemize}
        \item All the $\hat{\alpha}$ values for are the cases are above 95\%.
        \item The highest accuracy of 99.3\% is achieved with source code written in $Java$ programming languages.
        \item The accuracy values and F1 score values for all the languages are generally above 86\%.
    \end{itemize}
    \item \textit{Inference:} CRUSO-P performs equally well for all the programming languages. 
\end{itemize}

\subsection{Comparison of CRUSO-P with CRUSO}
\label{sec:comparison-of-CRUSO-P-with-CRUSO}

One of this work's key objectives is \textit{to significantly improve our previous work's \cite{Sodhi2019UsingSC} speed and efficiency.} When dealing with code reviews, a significant problem is the amount of time spent  performing them \cite{codereview2019wastetime,smartbear2019survey}. A code review assisting tool that provides accurate estimates but takes very long to deliver them will be practical of minimal use. Therefore, in this experiment, we evaluate the performance of our CRUSO-P in terms of \emph{response time} and \emph{memory usage}, with its previous version CRUSO.   

\textbf{A short recap of our previous Code Review Assisting tool --CRUSO \cite{Sodhi2019UsingSC}:} For an input source code $c$, CRUSO uses the Winnowing algorithm to identify SO posts containing code fragments similar to $c$, and analyzes the content of these relevant posts to estimate the defectiveness of $c$. CRUSO-P, in comparison to CRUSO, replaces the Winnowing algorithm with PVA. With PVA, the source code representation changes from the variable-length fingerprints to the fixed-length vectors. This experiment intends to investigate how this change in source code representation affects our tool's performance.

\textbf{The research question addressed:}
\textit{How does CRUSO-P perform in comparison to CRUSO? }

\subsubsection{Test-bed setup}
To perform this experiment, we implemented the existing approach \cite{Sodhi2019UsingSC} for the SO posts containing source code written in the considered set of programming languages, viz., C, C\#, Java, Python, and JavaScript.  To compare these tools' performance, we selected a random sample of 50 source files for each of the considered programming languages from different GitHub repositories (discussed in \S\ref{sec:selecting-GitHub}). We performed this experiment with the help of a group of programmers involved in developing software projects. 

\subsubsection{Procedure} 
The key steps involve the following:
\begin{enumerate}
    \item Compute the source code fingerprints for all the code fragments present in the considered SO posts. We use the Winnowing algorithm to perform this step.
    \item The obtained fingerprints are populated as a database table named as \textit{winnow}.
    \item Compare the storage used by the \textit{winnow} table and the \textit{vectors} table of SOpostsDB.
    \item \label{step:test-response-time}Compare the response time of CRUSO and CRUSO-P on testing with the selected random samples. 
\end{enumerate}

\subsubsection{Results and Observations} 
The salient observations are:
\begin{itemize}
        \item CRUSO-P has an average response time of 6.28 seconds, while the prior one based on Winnowing has 287.92 seconds. 
        \item The vectors table for CRUSO-P occupies 121.53 MBs, while the CRUSO's Winnowing table occupies of 14239 MBs.
\end{itemize}  

\textbf{Inference:} CRUSO-P achieves a speed improvement of 97.82\%  and a storage reduction of 99.15\% over CRUSO. The highest accuracy achieved by CRUSO-P is 99.3\% and 94\% in CRUSO \cite{Sodhi2019UsingSC}. Therefore, CRUSO-P achieves an improvement of 5.6\% in terms of accuracy when compared with CRUSO.

\subsection{Threats to validity}
\label{sec:threats-to-validity}
As observed from the experiments, the PVA models' accuracy depends on the training data's nature.  Thus the performance of the tool might vary if trained on a different dataset. The PVA models are trained on the source files written in languages \{C, C\#, Java, Python, JavaScript\}. Therefore, CRUSO-P can detect the defectiveness of the source files written in these programming languages only. However, we can extend this approach to other languages as well.

Further, while designing the function extraction interface based on ANTLR, we could not find the ANTLR grammars of C\# and JavaScript. Therefore, C\# and JavaScript, the \emph{function extractor} passes the input source code content to CRUSO-P for source code matching. 

While performing source code matching to detect the relevant SO posts, we considered the code fragments with the length $>=$ 100 characters (excluding the white spaces). We assume that the source code below this length would not represent any proper functionality, which also helps remove outliers from our dataset and remove the dataset's swamping effect \cite{chiang2007masking}.

A SO post generally comprises of multiple code fragments surrounding by text descriptions. One of the parameters that we use to infer the source code's defectiveness in SO is the text description present in the post.  Thus, there arises a need for mapping the code fragments with the constituent text fragments in various SO posts. While implementing this mapping procedure, we assume that \emph{A text description preceding a code snippet $c$ describes the nature of $c$.} The SO posts not adhering to this structuring of text and code fragments might result in false positives. 

Further, to compute the final defectiveness estimate, we use the \emph{majority vote principle} over the matching code-fragments' defectiveness estimates.  However, in the case of safety-critical software, there exists merit in being conservative. In that case, instead of the majority vote principle, it is safer to report the source code is \textit{likely to be defective} if there exists even a single defective code match.

\section{Conclusion}
\label{sec:conclusion}

Code review is an essential software quality assurance activity, intended to find software defects and to estimate the software quality. The existing code review methods are slow and inefficient. We present a novel tool -- CRUSO-P, which acts as a code review assistant for a programmer and helps in augmenting code reviews based on the information collected from SO posts. CRUSO-P works by determining the code similarity between the SO code fragments and the source code submitted as input. CRUSO-P leverages the PVA vector representations of source code present in SO posts to perform the code matching, thereby achieving an improvement of 97.82\% in response time and a storage reduction of 99.15\%, over one of the SOA tools. CRUSO-P achieves the best accuracy of 99.6\% in case of models trained on the C programming language. CRUSO-P and the vectors database can be used for building software tools in related application areas such as defectiveness estimation, code review and recommendation. Our results show that CRUSO-P outperforms the existing methods based on Winnowing algorithm and source code fingerprints.

\bibliographystyle{ACM-Reference-Format}
\bibliography{bibFile}
\end{document}